\documentclass[journal,twocolumn]{IEEEtran}

  \usepackage[pdftex]{graphicx}
  \graphicspath{{pic/}}

\usepackage[cmex10]{amsmath}
\usepackage[cmintegrals]{newtxmath}
\usepackage{mathrsfs}

\usepackage{pgfplots}
\usepackage{tikz}

\ifCLASSOPTIONcompsoc
 \usepackage[caption=false,font=normalsize,labelfont=sf,textfont=sf]{subfig}
\else
 \usepackage[caption=false,font=footnotesize]{subfig}
\fi
\usepackage{url}

\usepackage{cite}

\usepackage{multirow}
\usepackage{array}
\newcolumntype{L}[1]{>{\raggedright\let\newline\\\arraybackslash\hspace{0pt}}m{#1}}
\newcolumntype{C}[1]{>{\centering\let\newline\\\arraybackslash\hspace{0pt}}m{#1}}
\newcolumntype{R}[1]{>{\raggedleft\let\newline\\\arraybackslash\hspace{0pt}}m{#1}}

\usepackage{siunitx}

\newcommand{\erfc}{\mathrm{erfc}}
\newcommand{\binomial}[2]{\mathcal{B}\left( #1 , #2 \right)}

\usepackage{todonotes}
\definecolor{morange}{rgb}{0.8,0.2,0}
\definecolor{mblue}{rgb}{0,0.3,1.0}
\definecolor{mgreen}{rgb}{0.2,0.4,0}

\newcommand{\rxi}[1]{\text{Rx}_{#1}}
\newcommand{\txi}[1]{\text{Tx}_{#1}}
\newcommand{\ntx}{N_\text{Tx}}

\newcommand{\txt}{2\!\times\! 2}

\newcommand{\Ft}[1]{F(#1)}

\newcommand{\Fij}[1]{F_\text{#1}}

\newcommand{\uvec}{\mathbf{u}}
\newcommand{\uk}[1]{u[#1]}
\newcommand{\uik}[2]{u_{#1}[#2]}

\newcommand{\xik}[2]{x_{#1}[#2]}

\newcommand{\svec}{\mathbf{s}}

\newcommand{\sk}[1]{s_{#1}}
\newcommand{\yk}[1]{y[#1]}
\newcommand{\yik}[2]{y_{#1}[#2]}
\newcommand{\ycombk}[2]{y_\text{#1}[#2]}
\newcommand{\nk}[1]{n[#1]}
\newcommand{\nik}[2]{n_{#1}[#2]}
\newcommand{\uhatk}[1]{\hat{u}[#1]}
\newcommand{\utildek}[1]{\tilde{u}[#1]}

\newcommand{\hjil}[3]{h_{#1#2}[#3]}

\newcommand{\hl}[1]{h[#1]}
\newcommand{\hhatl}[1]{\hat{h}[#1]}

\newcommand{\G}{\mathbf{G}}
\newcommand{\I}{\mathbf{I}}
\newcommand{\hermit}{\mathrm{H}}

\newcommand{\N}{\mathbf{N}}

\newcommand{\Ts}{T_\text{s}}

\newcommand{\BER}{\mathrm{BER}}

\newcommand{\xth}[1]{{#1\text{th}}}

\begin{document}
\title{Spatial Diversity in Molecular Communications}
%
\author{Martin~Damrath,~\IEEEmembership{Student Member,~IEEE}, H. Birkan Yilmaz,~\IEEEmembership{Member,~IEEE},\\ Chan-Byoung Chae,~\IEEEmembership{Senior Member,~IEEE,} and Peter~Adam~Hoeher,~\IEEEmembership{Fellow,~IEEE}%
\thanks{M.~Damrath and P.\,A.~Hoeher are with the Faculty of Engineering, 
Kiel University, Germany, e-mail: \{md, ph\}@tf.uni-kiel.de.}%
\thanks{H.\,B.~Yilmaz and C.-B. Chae are with the Yonsei Institute of Convergence Technology, School of Integrated Technology, Yonsei University, Korea, e-mail: \{birkan.yilmaz, cbchae\}@yonsei.ac.kr.}
}

\markboth{}%
{Submitted paper}

\IEEEpubid{}

\maketitle

\begin{abstract}
In this work, spatial diversity techniques in the area of multiple-input multiple-output (MIMO) diffusion-based molecular communications (DBMC) are investigated. For transmitter-side spatial coding, Alamouti-type coding and repetition MIMO coding are proposed and analyzed. At the receiver-side, selection diversity, equal-gain combining, and maximum-ratio combining are studied as combining strategies. Throughout the numerical analysis, a symmetrical $\txt$ MIMO-DBMC system is assumed. Furthermore, a trained artificial neural network is utilized to acquire the channel impulse responses. The numerical analysis demonstrates that it is possible to achieve a diversity gain in molecular communications. In addition, it is shown that for MIMO-DBMC systems repetition MIMO coding is superior to Alamouti-type coding.
\end{abstract}


\IEEEpeerreviewmaketitle

\section{Introduction}
\IEEEPARstart{M}{olecular} communication (MC), a biologically inspired communication paradigm,
utilizes molecules as information carriers \cite{farsad2016comprehensiveSO,Akyildiz2008}.
MC is claimed to be a key technology in realizing autonomous nanomachines (NMs) \cite{Atakan2012}, the size of which ranges from several nanometers up to a few micrometers \cite{Xia2003}.
Due to their size, NMs are restricted with respect to their energy budget and capabilities \cite{Nakano2011,Nakano2013},
while MC provides an energy-efficient biocompatible method of communication.
Consequently, the capability of NMs can be enhanced by working as a swarm \cite{Akyildiz2008,guo2016molecularCC}.
MC can be used in the industrial and consumer sectors, such as with food and water quality control or intelligent textile fabrics.
MC can also be utilized in the environmental field, such as with biodegradation or air pollution control.
The main application, however, is anticipated to be in the medical sector,
where NMs can be used for applications like targeted drug delivery, tissue engineering, or health monitoring \cite{Atakan2012,Nakano2013}.

Diffusion-based molecular communication (DBMC) \cite{Pierobon2010} is a passive form of MC.
Following the law of diffusion, messenger molecules propagate passively from a source to a sink.
This offers an energy efficient way of communication,
because the energy for propagation comes directly from the environment.
However, the communication channel is fundamentally different from the classical radio-based wireless communication channel.
In fact, radio waves propagate deterministically in a given environment, whereas molecules perform a random walk. 
As a result, the diffusive propagation channel possess a slowly decreasing stochastic channel impulse response.
Consequently, DBMC systems suffer from intersymbol interference (ISI) and unreliable transmission~\cite{noel2014optimalRD}.
In multiple-input multiple-output (MIMO) scenarios, link reliability can be improved by spatial diversity through multiple transmit and/or receive antennas.

In classical wireless communications, MIMO techniques already appertain to the state-of-the-art.
In molecular communication, however, they have just rarely been considered.
To the best of our knowledge, the first conjunction between MC and MIMO was given in \cite{Meng2012a}.
The authors introduced transmitter diversity, receiver-side diversity combining, and spatial multiplexing to the area of DBMC.
While focusing on multi-user interference, the effect of ISI was paid little attention throughout the work.
In \cite{Koo2016}, several detection algorithms are proposed for spatial multiplexing scenarios in DBMC.
In contrast to \cite{Meng2012a}, the authors in \cite{Koo2016} took both, ISI and interlink interference (ILI), in their channel model into account.
Furthermore, they extended their tabletop molecular single-input single-output (SISO) testbed to a MIMO testbed.
The testbed was used to demonstrate an improvement with respect to the data rate when applying spatial multiplexing.
The authors in \cite{Lu2016} expanded a MC broadcast system by a second absorbing receiver and studied the effect on the bit error ratio (BER) and the channel capacity.

The focus of this paper is on a DBMC MIMO channel taking ISI as well as ILI into account.
The main contribution is a study of different spatial diversity algorithms at the transmitter and at the receiver sides.
At the transmitter side, we propose and analyze two different spatial coding techniques, namely Alamouti-type coding and repetition MIMO coding.
At the receiver side, we focus on three different receiver combining strategies: Selection diversity, equal-gain combining, and maximum-ratio combining.
In terms of BER simulations, we show the influence of key system parameters on the system performance.
Furthermore, we investigate the diversity gain compared to a SISO scenario.
Within the numerical simulations, we used a trained artificial neural network (ANN) to acquire the MIMO channel impulse responses.

The remainder of this paper is organized as follows: 
Section~\ref{Sec:SystemModel} summarizes a $\txt$ MIMO system model that is assumed throughout this work.
Section~\ref{Sec:ANN} presents how MIMO channel impulse responses are acquired by a trained ANN.
Section~\ref{Sec:SpatialDiversity} proposes, based on the system model, spatial coding techniques, as well as receiver combining strategies for molecular MIMO systems.
Section~\ref{Sec:DetectionAlgorithms} presents the detection algorithms that are applied in the numerical analysis in Section~\ref{Sec:NumericalResults}.
When necessary, the detection algorithms are adapted to the $\txt$ MIMO scenario.
Finally, Section~\ref{Sec:Conclusion} summarizes the work and gives an outlook for future work.

\IEEEpubidadjcol

\section{System Model\label{Sec:SystemModel}}

\subsection{Topology and Propagation Model}
\begin{figure} 
 \centering 
 \def\svgwidth{\columnwidth+0.5cm}
  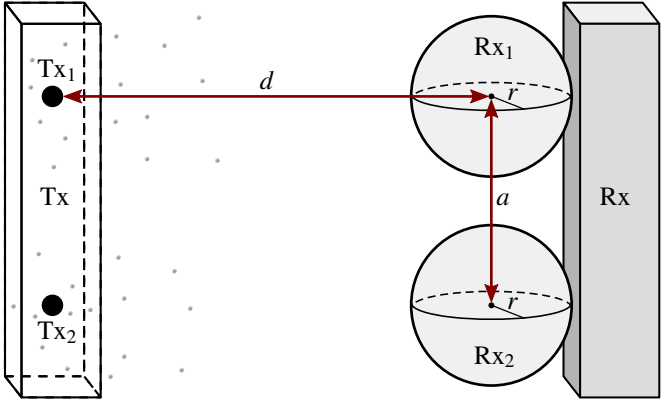
 \caption{Model of the diffusion-based molecular $2\times2$ MIMO system under investigation \cite{Koo2016}.}
 \label{Fig:SystemModel}
\end{figure}
The system model under investigation, shown in Fig.~\ref{Fig:SystemModel}, is similar to the system model introduced in \cite{Koo2016} and \cite{lee2015molecularMC}.
It contains a static transmitter Tx and a static receiver Rx within a fluid medium.
The Rx has two spherical receive apertures $\rxi{1}$ and $\rxi{2}$ with radius $r$ attached to its reflecting body.
The Tx consists of two point-emitters $\txi{1}$ and $\txi{2}$.
Subsequently, the receive apertures and the emitters are called antennas.
$\txi{1}$ and $\txi{2}$ are directly aligned to $\rxi{1}$ and $\rxi{2}$.
Consequently, the distance between $\txi{1}$ and $\rxi{1}$, as well as that between $\txi{2}$ and $\rxi{2}$, is given as $d$.
Furthermore, the separation distance between $\txi{1}$ and $\txi{2}$, as well as between $\rxi{1}$ and $\rxi{2}$, is given as $a$.
As a result, Tx and Rx form a symmetrical $\txt$~MIMO system.
The fluid medium is assumed to be homogeneous, three-dimensional infinitely spatially extended, and has no drift.
Accordingly, it is described by the constant diffusion coefficient $D$.

The molecules emitted by $\txi{1}$ and $\txi{2}$ propagate by Brownian motion, which is described by the Wiener process~\cite{srinivas2012molecular}. 
The Wiener process $W(t)$ is characterized as follows:
\begin{itemize}
\item $W(0) = 0$,
\item $W(t)$ is almost surely continuous, 
\item $W(t)$ has independent increments,
\item $W(t_2) - W(t_1) \sim \mathcal{N}(0,\,\phi(t_2-t_1))$ for $0\leq t_1 \leq t_2$,
\end{itemize}
where $\mathcal{N}(\mu, \sigma^2)$ is the Gaussian distribution with mean $\mu$ and variance $\sigma^2$. Simulating the Brownian motion includes consecutive steps in an $n$-dimensional space that obeys to Wiener process dynamics. For an accurate simulation, time is divided into sufficiently small time intervals ($\Delta t$) and at each time interval molecules take random steps in each dimension.
In an $n$-dimensional space, a random step is given as:
\begin{align}
\begin{split}
\Delta r &= (\Delta r_1, ..., \Delta r_n) \\
\Delta r_i &\sim \mathcal{N}(0, \, 2D\Delta t) \;\; \forall i \in \{1,..,n\},
\end{split}
\end{align}
where $\Delta r$ and $\Delta r_i$ correspond to random displacement vector and displacement at the $\xth{i}$ dimension. 

$\rxi{1}$ and $\rxi{2}$ are assumed to be perfect absorbing and counting receivers.
Accordingly, a diffusing molecule will be counted and removed from the environment the first time it hits to a receiving sphere.
As a result, the time histogram of absorbed molecules at $\rxi{1}$ and $\rxi{2}$ follow the first passage time concept.
Assuming just a single absorbing spherical receiver inside a 3-dimensional (3-D) environment,
the probability that a molecule hits the receiver until time $t$ after its release is given as \cite{Yilmaz2014}:
\begin{equation}\label{Eq:Fhit}
 \Ft{t} = \frac{r}{d} \ \erfc \left( \frac{d-r}{\sqrt{4Dt}} \right),
\end{equation}
where $\erfc(\cdot)$ denotes the complementary error function.
For multiple absorbing spherical receivers, unfortunately, there do not exists an equivalent closed-form expression.
Consequently, for a given MIMO scenario (\ref{Eq:Fhit}) has to be obtained by a random-walk-based simulation.
Alternatively, it can be acquired by using a trained ANN, as presented in Section~\ref{Sec:ANN}.

\subsection{Communication Channel}
The modulation scheme under investigation is on-off keying (OOK)~\cite{kuran2011modulationTF,yilmaz2014simulationSO,kim2013novel}. 
$\txi{i}$ emits either no molecules or $N$ messenger molecules at the beginning of a symbol period of length $\Ts$ to represent bit $\uik{i}{k}=0$ or $\uik{i}{k}=1$, respectively.
Accordingly, molecules emitted by $\txi{1}$ and $\txi{2}$ are of the same type.
They are assumed to be the only molecules in the medium; i.e., there is no background noise caused by molecules that are initially present in the medium.
Furthermore, Rx is assumed to be synchronized with Tx in time domain.
As an example, time synchronization can be achieved by an external signal like the human heart-beat or, as suggested in \cite{Moore2013}, by releasing inhibitory molecules.
In addition, $\rxi{1}$ and $\rxi{2}$ perform strength\slash energy detection \cite{Mahfuz2010a,Mahfuz2010b,Llatser2013}.
Consequently, the number of hitting molecules at each receive antenna is accumulated for each bit period separately.

The MIMO channel can be represented by a superposition of all subchannels.
Here, a subchannel is defined as the channel between the transmit antenna $\txi{i}$ and the receive antenna $\rxi{j}$.
Accordingly, each subchannel can be characterized by the corresponding channel coefficients $\hjil{j}{i}{\ell}$ ($0 \leq \ell \leq L$)
and can be represented by an equivalent discrete-time channel model with the effective channel memory length $L$ \cite{Damrath2017,genc2016isiAM}.
As a result, superimposing all subchannels related to $\rxi{j}$ will lead to the total number of received molecules at $\rxi{j}$:
\begin{equation}\label{Eq:ReceivedMolecules}
  \yik{j}{k} = \sum \limits_{i=1}^{\ntx} \sum \limits_{\ell=0}^L \hjil{j}{i}{\ell} \xik{i}{k\!-\!\ell} + \nik{j}{k},
\end{equation}
where $\ntx$ is the total number of transmitters,
$\hjil{j}{i}{\ell}$ describes the probability that a molecule hits $\rxi{j}$ during the $\ell$th time slot after its emission at $\txi{i}$,
$\xik{i}{k}$ is the discrete-time representation of the modulated data symbol transmitted by $\txi{i}$ at the start of the $\xth{k}$ transmission interval,
and $\nik{j}{k}$ describes the amplitude dependent noise caused by the diffusive propagation of the molecules.
For OOK, which is assumed throughout this paper, $\xik{i}{k}$ is defined as
\begin{equation} \label{Eq:Mapping}
  \xik{i}{k}  = \begin{cases}
            N &\quad \textnormal{if} \ \uik{i}{k}=1 \\
            0 &\quad \textnormal{if} \ \uik{i}{k}=0.
           \end{cases}
 \end{equation}
The event that a single molecule emitted by $\txi{i}$ is absorbed by $\rxi{j}$ during a certain time period can be modeled by a Bernoulli trial with success probability $\hjil{j}{i}{\ell}$.
Accordingly, the absorption event of $N$ molecules can be described by a binomial distribution~\cite{yilmaz2014arrivalMF}.
As a result, the distribution of (\ref{Eq:ReceivedMolecules}) follows the sum of several binomial distributions:
\begin{equation}
 \yik{j}{k} \sim \sum \limits_{i=1}^{\ntx} \sum \limits_{\ell=0}^L \binomial{\xik{i}{k\!-\!\ell}}{\hjil{j}{i}{\ell}},
\end{equation}
where $\binomial{M}{p}$ describe a binomial distribution with $M$ number of trials and success probability $p$.

Assuming a SISO scenario, the channel coefficients can be easily determined from (\ref{Eq:Fhit}):
\begin{equation}\label{Eq:hSISO}
 \hl{\ell} = \Ft{(\ell+1)\Ts} - \Ft{\ell\Ts}.
\end{equation}
If there is more than one absorbing sphere present inside the medium,
$\hjil{j}{i}{\ell}$ have to be determined differently.
One method includes to run random-walk-based simulations,
another method is to utilize a trained ANN as presented in Section~\ref{Sec:ANN}.

\section{ANN for Channel Modeling\label{Sec:ANN}}
We use the trained ANN from our previous work work~\cite{lee2017machineLB_ARXIV} to model a molecular MIMO channel. A trained ANN is able to estimate the channel coefficients $\hjil{j}{i}{\ell}$ for a given MIMO scenario. Please note that a trained ANN does not require any simulation data while we need simulations for the training phase. For training the ANN, we did extensive simulations and utilized a modified SISO channel response function for fitting to the simulation data and then we trained an ANN to predict the modified SISO channel function parameters. 

In a $\txt$ molecular MIMO scenario, we have two different spherical absorbing receivers -- $\rxi{1}$ and $\rxi{2}$ -- so that we need to model, for each receive antenna, two different channel impulse response functions per receive antenna, which depend on the distances.

We consider a case in which only $\txi{1}$ emits molecules for analyzing the cumulative channel impulse response functions at $\rxi{1}$ (i.e., $\Fij{11}(\cdot)$) and at $\rxi{2}$ (i.e., $\Fij{21}(\cdot)$) for modeling the received signal. Due to the rectangular symmetry, formulating $\Fij{11}(\cdot)$ and $\Fij{21}(\cdot)$ enables us to obtain $\Fij{22}(\cdot)=\Fij{11}(\cdot)$ and $\Fij{12}(\cdot)=\Fij{21}(\cdot)$. The modified channel impulse response function at $\rxi{1}$ is given as follows: 
\begin{align}
\begin{split}
\Fij{11} (t, b_1, b_2, b_3) =  b_1 \, \frac{r}{d} \,\erfc \left( \frac{d\!-\!r}{(4D)^{b_2} \, t^{b_3}}\right),
\end{split}
\label{eqn_model_rx1}
\end{align}
where $b_1$, $b_2$, and $b_3$ represent the model fitting parameters. These model-fitting parameters are introduced so as to compensate for the discrepancy between the SISO and MIMO models. Similarly we define the response at $\rxi{2}$ (due to the cross link interference) as follows:
\begin{align}
 \Fij{21} (t, b_4, b_5, b_6) =  b_4 \, \frac{r}{\sqrt{d^2\!+\!a^2}} \,\erfc \left( \frac{\sqrt{d^2\!+\!a^2}-r}{(4D)^{b_5} \, t^{b_6}}\right),
\end{align}
where $b_4$, $b_5$, and $b_6$ are model fitting parameters. 

\begin{figure}[!t]
	\begin{center}
		\includegraphics[width=1.0\columnwidth,keepaspectratio]%
		{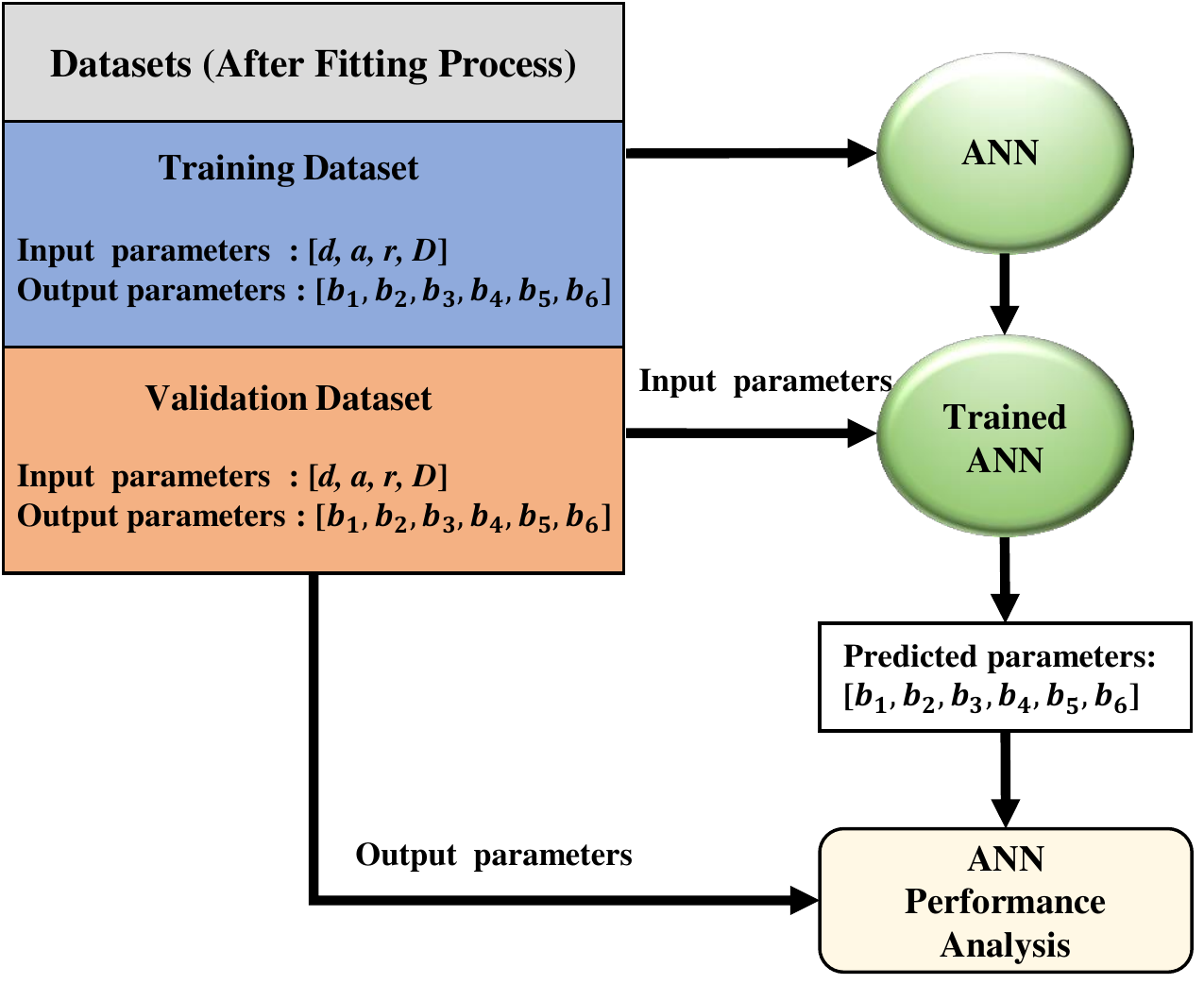}
		\caption{The flowchart and the dataset structure of the ANN training process. After curve fitting, the input-output pairs are fed to the training process, where the input is ${(d,\,h,\,R,\,D)}$ and the output is the model parameters (i.e., $b_i$'s). }
		\label{fig_ann_flowchart}
	\end{center}
\end{figure}
To find the $b_i$ values, we use a nonlinear least squares curve-fitting technique on the simulation data. These values are the basis of the training and test datasets with the scenario parameters (Fig.~\ref{fig_ann_flowchart}). Hence, the output of the curve-fitting process consists of the model parameters for each specific scenario. After forming the training and test datasets, the training data is fed into the ANN training process. Note that the trained ANN only requires the system parameters such as $d$, $a$, $r$, and $D$ (no simulation data).

\begin{figure}[!t]
	\begin{center}
    	\includegraphics[width=1.0\columnwidth,keepaspectratio]%
		{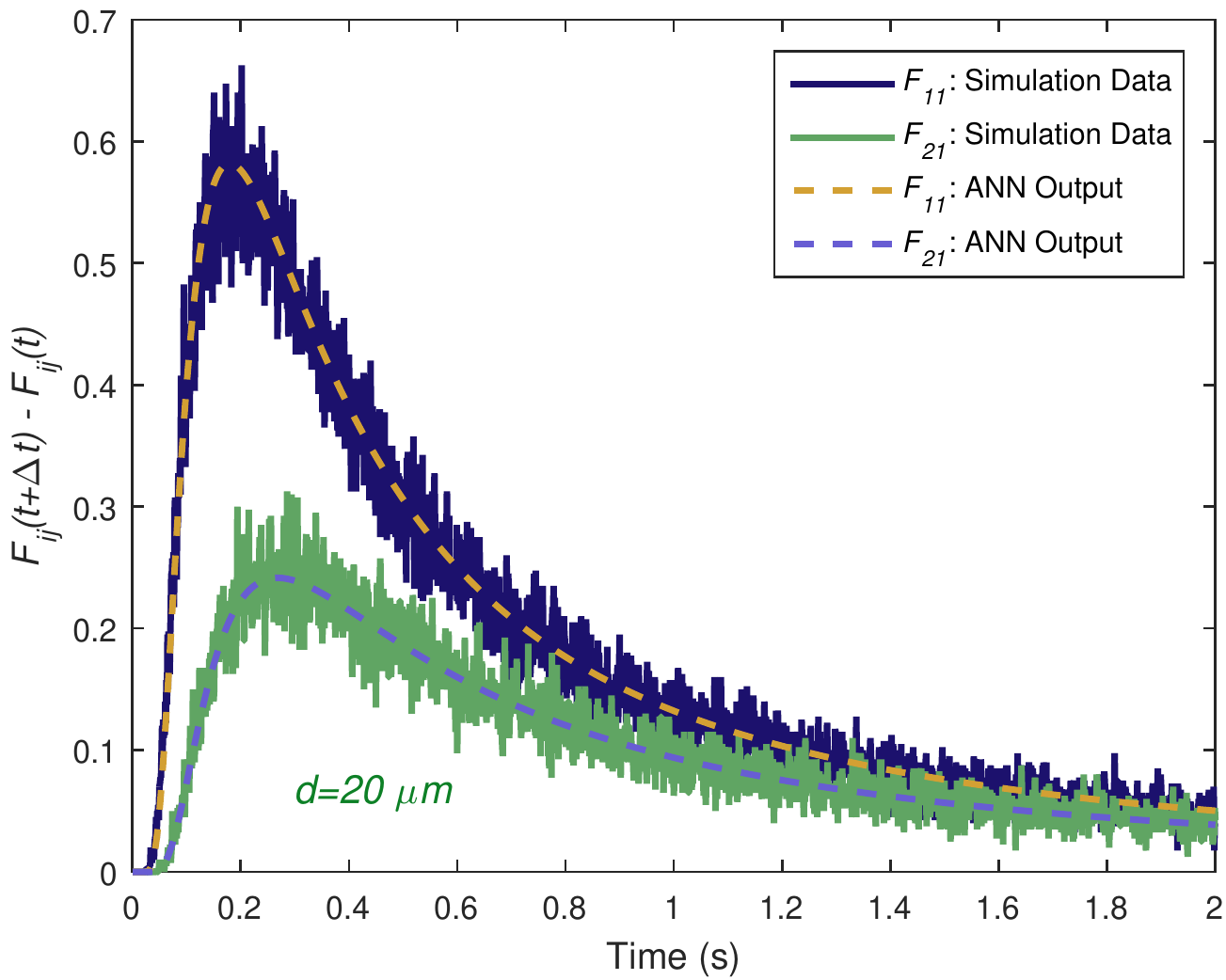}
    \end{center}
	\caption{ANN and simulation data comparison for channel impulse response functions after $\txi{1}$ emits 3000 molecules ($d\!=\!\SI{20}{\micro\meter}$, $a\!=\!\SI{13}{\micro\meter}$, $r\!=\!\SI{5}{\micro\meter}$, $D\!=\!\SI{200}{\micro\meter^2/\second}$, $\Delta t = \SI{0.001}{\second}$)}
	\label{fig_ann_vs_sim_res_deff15}
\end{figure}

\begin{figure}[!t]
	\begin{center}
    	\includegraphics[width=1.0\columnwidth,keepaspectratio]%
		{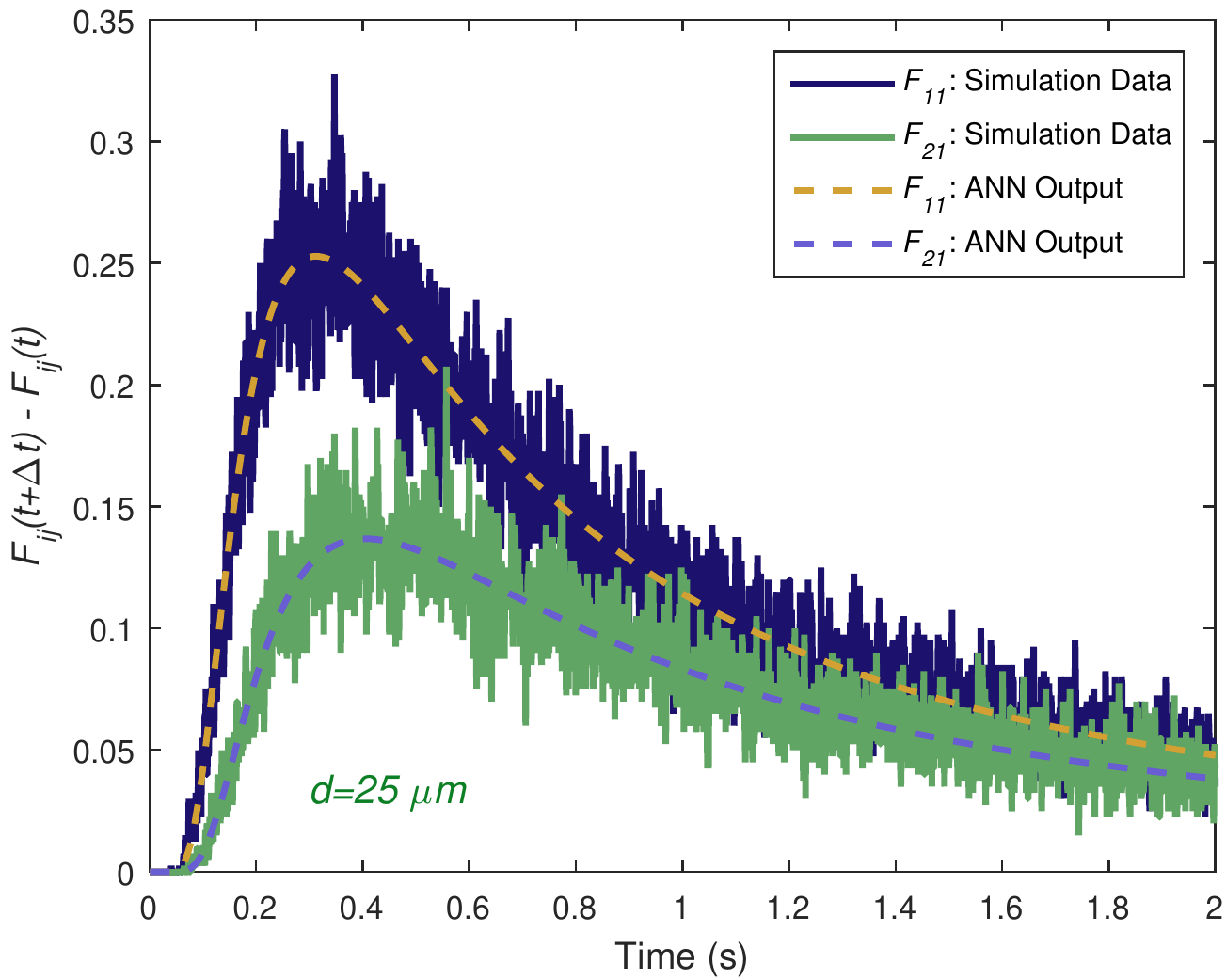}
    \end{center}
	\caption{ANN and simulation data comparison for channel impulse response functions after $\txi{1}$ emits 3000 molecules ($d\!=\!\SI{25}{\micro\meter}$, $a\!=\!\SI{13}{\micro\meter}$, $r\!=\!\SI{5}{\micro\meter}$, $D\!=\!\SI{200}{\micro\meter^2/\second}$, $\Delta t = \SI{0.001}{\second}$)}
	\label{fig_ann_vs_sim_res_deff20}
\end{figure}

In Figs.~\ref{fig_ann_vs_sim_res_deff15} and~\ref{fig_ann_vs_sim_res_deff20}, we present the resulting channel impulse functions with a time resolution of $\SI{0.001}{\second}$ for simulation and ANN technique. The received signal at the intended receiver (i.e., $\Fij{11}$ or $\Fij{22}$) and the ILI signal from the simulations are coherent with the ANN results. Therefore, we can utilize the output of the ANN to obtain channel coefficients for evaluating the number of received molecules and simulating the consecutive data transmissions. Depending on the symbol duration, we can evaluate the channel response for each symbol slot.

\begin{figure}[!t]
	\begin{center}
    	\includegraphics[width=1.0\columnwidth,keepaspectratio]%
		{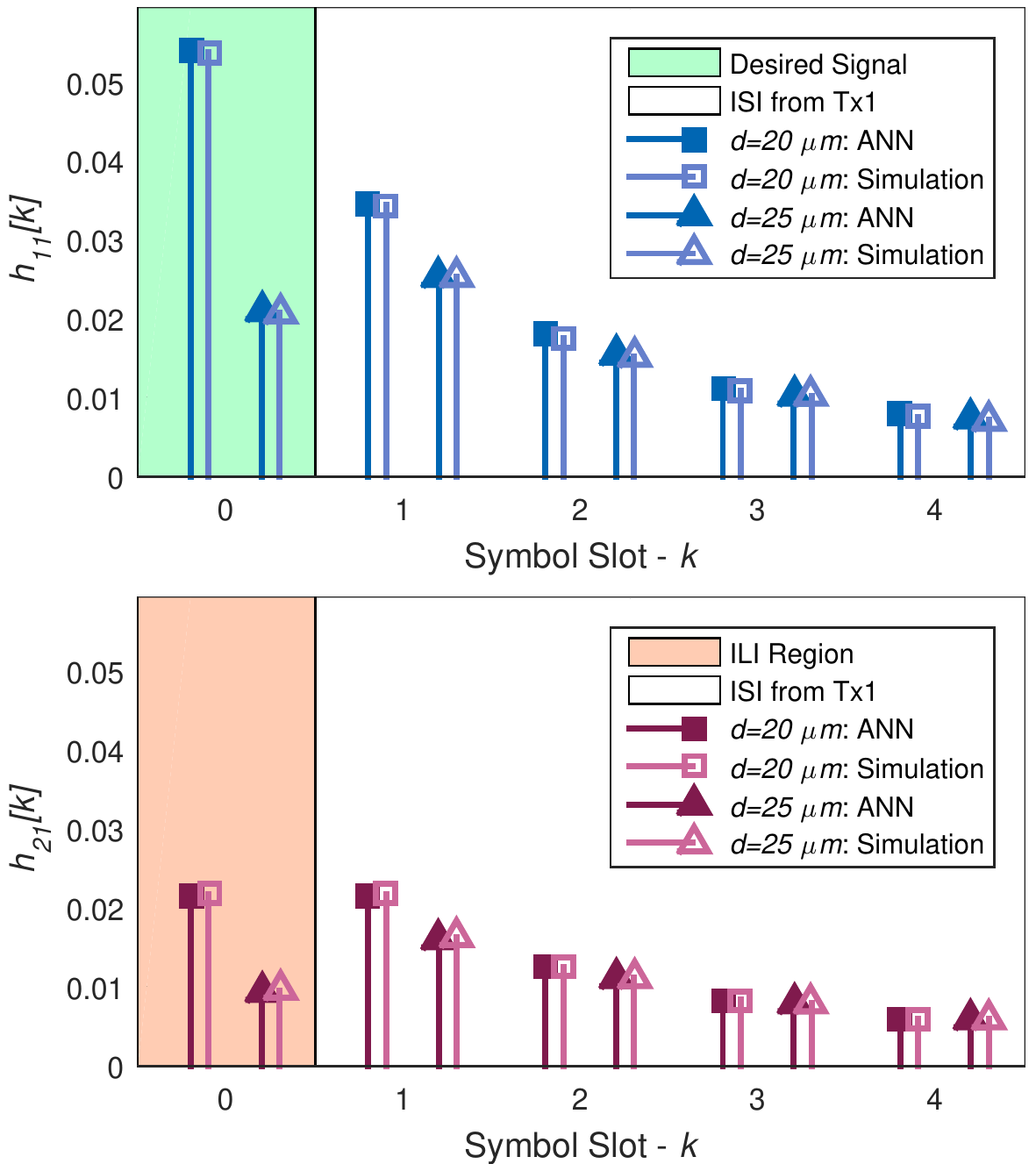}
    \end{center}
	\caption{Comparison of channel coefficients from ANN and simulation data for different distances ($a\!=\!\SI{13}{\micro\meter}$, $r\!=\!\SI{5}{\micro\meter}$, $D\!=\!\SI{200}{\micro\meter^2/\second}$, $\Ts = \SI{0.4}{\second}$)}
	\label{fig_ann_vs_sim_res_hijk}
\end{figure}

In Fig.~\ref{fig_ann_vs_sim_res_hijk}, we present the channel coefficients that are acquired from extensive simulations and the trained ANN. We plot the $\hjil{1}{1}{k}$ and $\hjil{2}{1}{k}$ values by utilizing $\Fij{11}$, $\Fij{21}$, and the symbol duration. The first observation is that the simulation and ANN results match well.
Our results validate and support the usage of ANN to obtain the channel coefficients.
Second, we observe that without equalization the symbol duration of $\SI{0.4}{\second}$ is not sufficient for $d\!=\!\SI{25}{\micro\meter}$. Thus, we clearly see the effect of distance on the channel coefficients while designing an MC system. For a $d\!=\!\SI{25}{\micro\meter}$ case (stems with triangle marker), the channel coefficient value at the current symbol slot is smaller than the first ISI symbol slot, which is also supported by Fig.~\ref{fig_ann_vs_sim_res_deff20} due to the peak time.
Therefore, with the help of the trained ANN, we are able to design a suitable symbol duration $\Ts$ for the cases of interest.

\section{Spatial Diversity\label{Sec:SpatialDiversity}}
In the case of multiple antennas at the transmitter and/or receiver side,
it is well-known from classical wireless communication that a spatial diversity gain can be achieved. This gain usually comes from spatial coding along multiple transmit antennas and/or receiver combining strategies given multiple receive antennas.
In the sense of spatial coding, the same information is transmitted over several antennas.
The information is typically represented by a sequence of data symbols $\svec$,
which is generated by mapping the binary data sequence $\uvec$ onto data symbols.
Below we present two different spatial coding techniques -- Alamouti-type coding and repetition MIMO coding.
For receiver-side combining strategies -- selection diversity, equal-gain combining, and maximum-ratio combining -- are suggested.

\subsection{Alamouti-type Coding}
The Alamouti scheme \cite{Alamouti1998} is an orthogonal space-time block code that was originally invented for two transmit antennas.
Its structure can be illustrated by the $\txt$ transmission matrix
\begin{equation}\label{Eq:Alamouti}
  \G = \begin{bmatrix}
                \sk{k} & \sk{k+1} \\ -\sk{k+1}^* & \sk{k}^*
                \end{bmatrix},
\end{equation}
where $\sk{k}$ denotes the $k$th data symbol of $\svec$.
The rows of $\G$ are related to the two consecutive transmission intervals $[k\Ts \ (k\!+\!1)\Ts]$ and $[(k\!+\!1)\Ts \ (k\!+\!2)\Ts]$, respectively.
The columns of $\G$ correspond to the two transmit antennas  $\txi{1}$ and $\txi{2}$, respectively.
Accordingly, in the first time slot, $\xik{1}{k}=\sk{k}$ is transmitted via $\txi{1}$ and $\xik{2}{k}=\sk{k+1}$ is transmitted simultaneously via $\txi{2}$.
In the second time slot, $\xik{1}{k\!+\!1}=-\sk{k+1}^*$ is transmitted via $\txi{1}$ and $\xik{2}{k\!+\!1}= \sk{k}^*$ is transmitted simultaneously via $\txi{2}$.
As a result, the information of both data symbols is spread over both transmit antennas, which provides a spatial diversity gain.
Note that $\G^\hermit\G=c\I$, where $\G^\hermit$ is the Hermitian of matrix $\G$, $c$ is a constant factor, and $\I$ denotes the identity matrix.
Consequently, the Alamouti scheme is an orthogonal space-time block code for complex data symbols.
With help of orthogonality, ILI can be canceled completely at the receiver side.
Thus, the realization of a maximum-likelihood detector can be simplified,
which makes Alamouti scheme popular in radio-based communication systems without ISI.

In the case of ISI, however, orthogonality is not maintained any more.
As a result, we must apply more complex detection algorithms such as maximum-likelihood sequence estimation (MLSE) \cite{Mietzner2004}.
Orthogonality can also be maintained by modifying the Alamouti transmission scheme, as shown in \cite{Lindskog2000}.
There is still a need, though, for equalization algorithms like MLSE.
Furthermore, the modification introduces additional errors at the edges of the proposed transmission blocks.

The Alamouti scheme assumes complex data symbols that can either be positive or negative.
In MC, however, the data symbols are typically represented by the amount of emitted molecules.
Thus, the data symbols are non-negative and real-valued.
Consequently, (\ref{Eq:Alamouti}) has to be modified to an Alamouti-type code \cite{Simon2005} that avoids minus signs and complex conjugation.
The focus of this work is on OOK.
Accordingly, following the principle of (\ref{Eq:Mapping}),
data bits are mapped onto data symbols $\sk{k}\in\{0,N\}$ \cite{kuran2011modulationTF,yilmaz2014simulationSO,kim2013novel}.
As suggested in \cite{Simon2005}, the adaptation to an Alamouti-type code can be done by the following two steps:
\begin{enumerate}
 \item Since there are only real-valued integer values, the complex conjugate operation can be discarded.
 \item The negative symbols can be replaced by the inverse symbol $\bar{s}_k:=N-\sk{k}$.
\end{enumerate}
Applying those two steps to (\ref{Eq:Alamouti}) leads to the transmission matrix of the Alamouti-type code:
\begin{equation}\label{Eq:a_G}
  \G = \begin{bmatrix}
                       \sk{k} & \sk{k+1} \\ N-\sk{k+1} & \sk{k}
                     \end{bmatrix}.
\end{equation}
In \cite{Simon2005}, a maximum likelihood detection metric was derived.
Note that in molecular communication, however, the system is affected by ISI and the orthogonality of the Alamouti-type code is no longer maintained.
Thus, more complex detection algorithms like MLSE have to be applied for detection.

\subsection{Repetition MIMO Coding}
Repetition MIMO coding \cite{Wilson2005} offers a simple intuitive alternative to orthogonal Alamouti scheme.
In contrast to the Alamouti scheme, the information is coded only in the spatial domain,
while the time domain is not exploited.
In detail, exactly the same data symbol is transmitted via each transmit antenna at the same time.
Accordingly for a $\txt$ MIMO scenario, the transmission matrix is defined as
\begin{equation}
 \G = \begin{bmatrix}
       \sk{k} & \sk{k}
      \end{bmatrix}.
\end{equation}
Note that there is no orthogonality in the code and ILI cannot be canceled out at the receiver side.
The ILI, however, will have a constructive influence of the signal strength.
As a result, even in the presence of ISI, SISO detection algorithms can be used at the receiver side.

\subsection{Receiver Combining}
If there is more than one receive antenna,
the received signals from each antenna have to be combined/selected,
before detection can be performed.
Normally, the selection/combining is done in one of three ways.
With selection diversity (SD), the strongest signal of all antennas is selected for detection.
In a molecular communication system in conjunction with OOK and ISI,
it is hard to determine which antenna receives the strongest signal.
For $\uk{k}=0$ the signal with the minimum number of received molecules would be the strongest one, 
while for $\uk{k}=1$ the signal with the maximum number of received molecules would be the strongest one.
In this work, a symmetrical scenario is considered.
Hence, the expected signal strength at both receive antennas is equal.
Therefore, without loss of generality, $\rxi{1}$ is selected in the case of SD:
\begin{align}
\begin{split}
 \ycombk{SD}{k} &=\yik{1}{k} \\ &=\sum \limits_{\ell=0}^L \hjil{1}{1}{\ell}\xik{1}{k\!-\!\ell} + \sum \limits_{\ell=0}^L \hjil{1}{2}{\ell}\xik{2}{k\!-\!\ell} + \nik{1}{k}.
\end{split} 
\end{align}
Another combining strategy is equal-gain combining (EGC), 
where the signals of all receive antennas are equally weighted and combined.
Adjusting the weighting factors to the corresponding channel quality leads to maximum-ratio combining (MRC),
which is equal to a maximum-likelihood receiver.
Consequently, channel knowledge is necessary at the receiver side.
However, in the case of a symmetrical scenario, which leads to equal channels at both receive antennas,
EGC is equal to MRC.
As a result, EGC is considered in the following:
\begin{equation}\label{Eq:EGC}
 \ycombk{EGC}{k}=\yik{1}{k}+\yik{2}{k}.
\end{equation}
Due to the symmetrical system setting, 
the channel description for EGC can be further simplified.
Considering that $\hjil{1}{1}{\ell}=\hjil{2}{2}{\ell}$ and $\hjil{1}{2}{\ell}=\hjil{2}{1}{\ell}$,
(\ref{Eq:EGC}) can be restated as
\begin{align}\label{Eq:yEGC}
 \ycombk{EGC}{k}=\sum\limits_{\ell=0}^L \hl{\ell} \left( \xik{1}{k\!-\!\ell}+\xik{2}{k\!-\!\ell} \right) + \nk{k},
\end{align}
where $\hl{\ell}\doteq\hjil{1}{1}{\ell}+\hjil{1}{2}{\ell}$ and $\nk{k}\doteq\nik{1}{k}+\nik{2}{k}$.

\section{Detection Algorithms\label{Sec:DetectionAlgorithms}}
For the bit error analysis throughout this paper, we consider and adopt from \cite{Damrath2016} three different detection algorithms.
First of all, the common fixed threshold detector (FTD)
\begin{equation}
 \uhatk{k} =  \begin{cases}
          1 & \quad \text{if } \yk{k} > \eta\\
	  0 & \quad \text{if } \yk{k} \leq \eta\\
  \end{cases}
\end{equation}
is used, where the threshold $\eta$ is chosen to be optimal in terms of minimizing the BER.
The optimal threshold is determined by means of an exhaustive search.
Second, the low-complexity adaptive threshold detector (ATD) is applied:
\begin{equation}
 \uhatk{k} =  \begin{cases}
          1 & \quad \text{if } \yk{k} > \yk{k\!-\!1}\\
	  0 & \quad \text{if } \yk{k} \leq \yk{k\!-\!1}.\\
  \end{cases}
\end{equation}
Note that ATD does not need any channel knowledge and inherently benefits from ISI.
The third algorithm is maximum-likelihood sequence estimation.
It is applied with the suboptimal squared Euclidean distance branch metric
\begin{equation}
  \gamma(\yk{k}|\left[\utildek{k},\dots,\utildek{k\!-\!L}\right])=\left( \yk{k} - \sum \limits_{\ell=0}^L N \hhatl{\ell} \utildek{k\!-\!\ell} \right)^2.
\end{equation}
During the numerical analysis, it is assumed that $\hhatl{\ell}$ are equal to the channel coefficients utilized in the equivalent discrete-time channel model.
Depending on the considered spatial coding and receiver combining strategy $\hhatl{\ell}$ has to be adapted.
For a SISO system, $\hhatl{\ell}$ is set equal to the channel coefficients from (\ref{Eq:hSISO}).
For repetition coding it yields $\hhatl{\ell}=\hl{\ell}$ for SD and $\hhatl{\ell}=2\hl{\ell}$ for EGC,
where $\hl{\ell}$ is defined as in (\ref{Eq:yEGC}).

In Alamouti-type coding, the information of two symbols is spread over two consecutive time slots.
Thus, the branch metric can be evaluated jointly over both time slots.
As a result, the branch metric has to be further adapted.

As an example, we present an Alamouti-type $\txt$ MIMO transmission scenario for $L=1$ as follows:
\begin{table}[h]
\caption{Transmission example for Alamouti-type $\txt$ MIMO scenario.}
\centering
\begin{tabular}{L{2.5cm} l l l l}
 \hline 
 Discrete time step $k$ $\vphantom{\frac{1}{2}}$ & $k\!-\!2$ & $k\!-\!1$ & $k$ & $k\!+\!1$ \\  
 \hline 
 $\xik{1}{k}$   & $\sk{k-2}$  & $N-\sk{k-1}$  & $\sk{k}$  & $N-\sk{k+1}$ \\
 $\xik{2}{k}$   & $\sk{k-1}$  & $\sk{k-2}$  & $\sk{k+1}$  & $\sk{k}$ \\
 \hline
\end{tabular}
\label{Tab_example_scenario}
\end{table}

\noindent Therefore, we need to consider the transmission matrix $\G$ in (\ref{Eq:a_G}) including the corresponding ISI terms:
\begin{equation}
  \G_1 = \begin{bmatrix}
          \sk{k} & \sk{k+1} & N-\sk{k-1} & \sk{k-2} \\ 
          N-\sk{k+1} & \sk{k} & \sk{k} & \sk{k+1}
        \end{bmatrix},
\end{equation}
where $\G_L$ represents the transmission matrix $\G$ with $L$ ISI terms.
In our example, the last two columns correspond to the transmitted signal in the corresponding previous time slot by $\txi{1}$ and $\txi{2}$, respectively.
Note that the first row's previous slot is $k\!-\!1$ and the second row's previous slot is $k$. In this case, the number of received molecules can be written as follows:
\begin{align}
\begin{split}
	\begin{bmatrix}
     \yik{1}{k} & \yik{2}{k} \\ \yik{1}{k\!+\!1} & \yik{2}{k\!+\!1} 
    \end{bmatrix}
    = \G_1 
    \begin{bmatrix}
       \hjil{1}{1}{0} & \hjil{2}{1}{0} \\
       \hjil{1}{2}{0} & \hjil{2}{2}{0} \\
       \hjil{1}{1}{1} & \hjil{2}{1}{1} \\
       \hjil{1}{2}{1} & \hjil{2}{2}{1} 
    \end{bmatrix} 
    + \begin{bmatrix} \N_1 \N_2 \end{bmatrix}, 
\end{split}
\end{align}
where $\N_i=[\,\nik{i}{k} \;\,\nik{i}{k\!+\!1}\,]^\mathrm{T}$.
Assuming a symmetrical scenario and EGC, the branch metric can be adapted as shown in (\ref{Eq:a_MLSE}).
\begin{figure*}[ht!]
 \normalsize
\begin{align}\label{Eq:a_MLSE}
\begin{split}
 \gamma(\yk{k},\yk{k\!+\!1}|\utildek{k\!+\!1},\utildek{k},\utildek{k\!-\!1},\utildek{k\!-\!2}) =& 
 \left[ \yk{k} - N\hl{0} (\utildek{k} + \utildek{k\!+\!1}) - N\hl{1}(\utildek{k\!-\!2}-\utildek{k\!-\!1} + 1) \right]^2 \\
 +&\left[ \yk{k\!+\!1} - N\hl{0}(\utildek{k} - \utildek{k\!+\!1} + 1) - N\hl{1} (\utildek{k}+\utildek{k\!+\!1}) \right]^2 
\end{split}
\end{align}
 \hrulefill
 \vspace*{4pt}
\end{figure*}

\section{Numerical Results\label{Sec:NumericalResults}}
\begin{table}
\caption{Simulation parameters used for analysis. The default parameters are in bold face.}
\centering
\begin{tabular}{L{2.2cm} l}
 \hline $\vphantom{\frac{1}{2}}$Parameter & Value\\ \hline
 $N$ & $\{500,\mathbf{1\:\!000},1\:\!500,2\:\!000 \}$ \\ 
 $\Ts \ [\si{\second}]$ & $\{0.48, \mathbf{0.6}, 0.8, 1.2 \}$ \\ 
 $(L+1)\Ts \ [\si{\second}]$ & $2.4$ \\ 
 $d \ [\si{\micro\meter}]$ & $\{ 10, 15, \mathbf{20}, 25\}$ \\ 
 $a \ [\si{\micro\meter}]$ & $\{ \mathbf{11}, 13, 15, 17\}$ \\ 
 $D \ [\si{\micro\meter^2/\second}]$ & $\{ 50, \mathbf{100}, 150, 200 \}$ \\ 
 $r \ [\si{\micro\meter}]$ & $5$ \\ 
 $K$ & $10^6$ \\ 
 $R$ & $1000$ \\
 \hline
\end{tabular}
\label{Tab:Parameter}
\end{table}
In this section, results from the numerical analysis are presented.
To analyze the effect of spatial diversity, BER simulations are performed for SISO and $\txt$ MIMO scenarios.
In detail, the impact of number of molecules $N$, symbol duration $\Ts$, transmission distance $d$, separation distance $a$, and diffusion coefficient $D$ on the BER is shown.
The simulation parameters are summarized in Table~\ref{Tab:Parameter},
where $K$ is the bit sequence length for one channel realization and $R$ is the total number of channel realizations.
Throughout the simulations, it is assumed that the remaining ISI is negligible after $(L+1)\Ts=\SI{2.4}{\second}$.
Accordingly, the channel memory length $L\in\{1, 2, 3, 4\}$ varies with different $\Ts$.
In the SISO scenarios, there is just a single transmit and a single receive antenna in the environment.
Furthermore, we set the number of emitted molecules $N$ to be twice as large as that in the MIMO scenarios.
This guarantees a fair comparison between SISO and $\txt$ MIMO scenarios by means of transmitting energy.

\begin{figure*}
\begin{footnotesize}
\begin{tabular}{p{0.43\textwidth} p{0.43\textwidth}}
\subfloat[{Variation of numbers of molecules.}]{
\begin{tikzpicture}[/pgfplots/tick scale binop=\times]
\begin{semilogyaxis}
[width=\columnwidth, height=0.83\columnwidth, grid=both,
grid style={dotted}, /pgf/number format/1000 sep={\:\!},
xmin=500, xmax=2000, ymax=0.5, ymin=1e-7,
legend columns=1,
legend cell align=left,
legend to name=Leg:Legend,
xlabel={$N$}, ylabel={$\BER$}]
\addplot [gray, line width=1, only marks, mark=square*] coordinates{(10,1) (20,1)};
\addplot [gray, line width=1, only marks, mark=triangle*] coordinates{(10,1) (20,1)};
\addplot [gray, line width=1, only marks, mark=diamond*] coordinates{(10,1) (20,1)};
\addplot [black, line width=1.5, densely dashed] coordinates{(10,1) (20,1)};
\addplot [red!60!black, line width=1.5, densely dashdotted] coordinates{(10,1) (20,1)};
\addplot [blue!60!black, line width=1.5] coordinates{(10,1) (20,1)};
\addplot [green!40!black, line width=1.5, densely dotted] coordinates{(10,1) (20,1)};
\addplot [black, line width=1, mark=square*, mark repeat=1, mark options=solid, densely dashed] table[x=N, y=BER_FTD_SU] {./pic/varN.dat};
\addplot [blue!60!black, line width=1, mark=square*, mark repeat=1, mark options=solid] table[x=N, y=BER_r_FTD_EGC] {./pic/varN.dat};
\addplot [black, line width=1, mark=triangle*, mark repeat=1, mark options=solid, densely dashed] table[x=N, y=BER_ATD_SU] {./pic/varN.dat};
\addplot [red!60!black, line width=1, mark=triangle*, mark repeat=1, mark options=solid, dashdotted] table[x=N, y=BER_r_ATD_SD] {./pic/varN.dat};
\addplot [blue!60!black, line width=1, mark=triangle*, mark repeat=1, mark options=solid] table[x=N, y=BER_r_ATD_EGC] {./pic/varN.dat};
\addplot [black, line width=1, mark=diamond*, mark repeat=1, mark options=solid, densely dashed] table[x=N, y=BER_MLSE_SU] {./pic/varN.dat};
\addplot [red!60!black, line width=1, mark=diamond*, mark repeat=1, mark options=solid, dashdotted] table[x=N, y=BER_r_MLSE_SD] {./pic/varN.dat};
\addplot [blue!60!black, line width=1, mark=diamond*, mark repeat=1, mark options=solid] table[x=N, y=BER_r_MLSE_EGC] {./pic/varN.dat};
\addplot [green!40!black, line width=1, mark=diamond*, mark repeat=1, mark options=solid, densely dotted] table[x=N, y=BER_a_MLSE_EGC] {./pic/varN.dat};
\legend{\small{Fixed threshold detector (FTD)}, \small{Adaptive threshold detector (ATD)}, \small{MLSE}, \small{SISO}, \small{Repetition MIMO - SD}, \small{Repetition MIMO - EGC}, \small{Alamouti-type - EGC}}
\end{semilogyaxis}
\end{tikzpicture}
\label{Fig:BER_Nmol}}
&
\subfloat[{Variation of symbol duration.}]{
\begin{tikzpicture}[/pgfplots/tick scale binop=\times]
\begin{semilogyaxis}
[width=\columnwidth, height=0.83\columnwidth, grid=both,
grid style={dotted}, /pgf/number format/1000 sep={\:\!},
xmin=0.48, xmax=1.2, ymax=0.5, ymin=1e-7,
xlabel={$\Ts$ in $\si{\second}$}, ylabel={$\BER$}]
\addplot [black, line width=1, mark=square*, mark repeat=1, mark options=solid, densely dashed] table[x=T, y=BER_FTD_SU] {./pic/varT.dat};
\addplot [blue!60!black, line width=1, mark=square*, mark repeat=1, mark options=solid] table[x=T, y=BER_r_FTD_EGC] {./pic/varT.dat};
\addplot [black, line width=1, mark=triangle*, mark repeat=1, mark options=solid, densely dashed] table[x=T, y=BER_ATD_SU] {./pic/varT.dat};
\addplot [red!60!black, line width=1, mark=triangle*, mark repeat=1, mark options=solid, dashdotted] table[x=T, y=BER_r_ATD_SD] {./pic/varT.dat};
\addplot [blue!60!black, line width=1, mark=triangle*, mark repeat=1, mark options=solid] table[x=T, y=BER_r_ATD_EGC] {./pic/varT.dat};
\addplot [black, line width=1, mark=diamond*, mark repeat=1, mark options=solid, densely dashed] table[x=T, y=BER_MLSE_SU] {./pic/varT.dat};
\addplot [red!60!black, line width=1, mark=diamond*, mark repeat=1, mark options=solid, dashdotted] table[x=T, y=BER_r_MLSE_SD] {./pic/varT.dat};
\addplot [blue!60!black, line width=1, mark=diamond*, mark repeat=1, mark options=solid] table[x=T, y=BER_r_MLSE_EGC] {./pic/varT.dat};
\addplot [green!40!black, line width=1, mark=diamond*, mark repeat=1, mark options=solid, densely dotted] table[x=T, y=BER_a_MLSE_EGC] {./pic/varT.dat};
\end{semilogyaxis}
\end{tikzpicture} 
\label{Fig:BER_Tb}}\\ 
\subfloat[{Variation of transmission distance.}]{
\begin{tikzpicture}[/pgfplots/tick scale binop=\times]
\begin{semilogyaxis}
[width=\columnwidth, height=0.83\columnwidth, grid=both,
grid style={dotted}, /pgf/number format/1000 sep={\:\!},
xmin=10, xmax=25, ymax=0.5, ymin=1e-7,
xlabel={$d$ in $\si{\micro\meter}$}, ylabel={$\BER$}]
\addplot [black, line width=1, mark=square*, mark repeat=1, mark options=solid, densely dashed] table[x=d, y=BER_FTD_SU] {./pic/varDist.dat};
\addplot [blue!60!black, line width=1, mark=square*, mark repeat=1, mark options=solid] table[x=d, y=BER_r_FTD_EGC] {./pic/varDist.dat};
\addplot [black, line width=1, mark=triangle*, mark repeat=1, mark options=solid, densely dashed] table[x=d, y=BER_ATD_SU] {./pic/varDist.dat};
\addplot [red!60!black, line width=1, mark=triangle*, mark repeat=1, mark options=solid, dashdotted] table[x=d, y=BER_r_ATD_SD] {./pic/varDist.dat};
\addplot [blue!60!black, line width=1, mark=triangle*, mark repeat=1, mark options=solid] table[x=d, y=BER_r_ATD_EGC] {./pic/varDist.dat};
\addplot [black, line width=1, mark=diamond*, mark repeat=1, mark options=solid, densely dashed] table[x=d, y=BER_MLSE_SU] {./pic/varDist.dat};
\addplot [red!60!black, line width=1, mark=diamond*, mark repeat=1, mark options=solid, dashdotted] table[x=d, y=BER_r_MLSE_SD] {./pic/varDist.dat};
\addplot [blue!60!black, line width=1, mark=diamond*, mark repeat=1, mark options=solid] table[x=d, y=BER_r_MLSE_EGC] {./pic/varDist.dat};
\addplot [green!40!black, line width=1, mark=diamond*, mark repeat=1, mark options=solid, densely dotted] table[x=d, y=BER_a_MLSE_EGC] {./pic/varDist.dat};
\end{semilogyaxis}
\end{tikzpicture}
\label{Fig:BER_d}}
&
\subfloat[{Variation of separation distance.}]{%
\begin{tikzpicture}[/pgfplots/tick scale binop=\times]
\begin{semilogyaxis}
[width=\columnwidth, height=0.83\columnwidth, grid=both,
grid style={dotted}, /pgf/number format/1000 sep={\:\!},
xmin=11, xmax=17, ymax=0.5, ymin=1e-7,
xlabel={$a$ in $\si{\micro\meter}$}, ylabel={$\BER$}]
\addplot [black, line width=1, mark=square*, mark repeat=1, mark options=solid, densely dashed] table[x=a, y=BER_FTD_SU] {./pic/varSep.dat};
\addplot [blue!60!black, line width=1, mark=square*, mark repeat=1, mark options=solid] table[x=a, y=BER_r_FTD_EGC] {./pic/varSep.dat};
\addplot [black, line width=1, mark=triangle*, mark repeat=1, mark options=solid, densely dashed] table[x=a, y=BER_ATD_SU] {./pic/varSep.dat};
\addplot [red!60!black, line width=1, mark=triangle*, mark repeat=1, mark options=solid, dashdotted] table[x=a, y=BER_r_ATD_SD] {./pic/varSep.dat};
\addplot [blue!60!black, line width=1, mark=triangle*, mark repeat=1, mark options=solid] table[x=a, y=BER_r_ATD_EGC] {./pic/varSep.dat};
\addplot [black, line width=1, mark=diamond*, mark repeat=1, mark options=solid, densely dashed] table[x=a, y=BER_MLSE_SU] {./pic/varSep.dat};
\addplot [red!60!black, line width=1, mark=diamond*, mark repeat=1, mark options=solid, dashdotted] table[x=a, y=BER_r_MLSE_SD] {./pic/varSep.dat};
\addplot [blue!60!black, line width=1, mark=diamond*, mark repeat=1, mark options=solid] table[x=a, y=BER_r_MLSE_EGC] {./pic/varSep.dat};
\addplot [green!40!black, line width=1, mark=diamond*, mark repeat=1, mark options=solid, densely dotted] table[x=a, y=BER_a_MLSE_EGC] {./pic/varSep.dat};
\end{semilogyaxis}
\end{tikzpicture}
\label{Fig:BER_a}}\\
\subfloat[{Variation of diffusion coefficient.}]{%
\begin{tikzpicture}[/pgfplots/tick scale binop=\times]
\begin{semilogyaxis}
[width=\columnwidth, height=0.83\columnwidth, grid=both,
grid style={dotted}, /pgf/number format/1000 sep={\:\!},
xmin=50, xmax=200, ymax=0.5, ymin=1e-7,
xlabel={$D$ in $\si{\micro\meter^2/\second}$}, ylabel={$\BER$}]
\addplot [black, line width=1, mark=square*, mark repeat=1, mark options=solid, densely dashed] table[x=D, y=BER_FTD_SU] {./pic/varD.dat};
\addplot [blue!60!black, line width=1, mark=square*, mark repeat=1, mark options=solid] table[x=D, y=BER_r_FTD_EGC] {./pic/varD.dat};
\addplot [black, line width=1, mark=triangle*, mark repeat=1, mark options=solid, densely dashed] table[x=D, y=BER_ATD_SU] {./pic/varD.dat};
\addplot [red!60!black, line width=1, mark=triangle*, mark repeat=1, mark options=solid, dashdotted] table[x=D, y=BER_r_ATD_SD] {./pic/varD.dat};
\addplot [blue!60!black, line width=1, mark=triangle*, mark repeat=1, mark options=solid] table[x=D, y=BER_r_ATD_EGC] {./pic/varD.dat};
\addplot [black, line width=1, mark=diamond*, mark repeat=1, mark options=solid, densely dashed] table[x=D, y=BER_MLSE_SU] {./pic/varD.dat};
\addplot [red!60!black, line width=1, mark=diamond*, mark repeat=1, mark options=solid, dashdotted] table[x=D, y=BER_r_MLSE_SD] {./pic/varD.dat};
\addplot [blue!60!black, line width=1, mark=diamond*, mark repeat=1, mark options=solid] table[x=D, y=BER_r_MLSE_EGC] {./pic/varD.dat};
\addplot [green!40!black, line width=1, mark=diamond*, mark repeat=1, mark options=solid, densely dotted] table[x=D, y=BER_a_MLSE_EGC] {./pic/varD.dat};
\end{semilogyaxis}
\end{tikzpicture}
\label{Fig:BER_D}}
&
\subfloat{%
\begin{tikzpicture}
\node at (0,0) {};
\node (leg) at (4.7,3.5) {\pgfplotslegendfromname{Leg:Legend}};
\end{tikzpicture}}
\end{tabular}
\end{footnotesize}
\caption{Bit error rate performance as a function of the number of molecules \protect\subref{Fig:BER_Nmol},
symbol duration \protect\subref{Fig:BER_Tb}, transmitting distance \protect\subref{Fig:BER_d}, separation distance \protect\subref{Fig:BER_a},
and diffusion coefficient \protect\subref{Fig:BER_D}.
If the corresponding parameter is not varying, it is fixed to $N=1\:\!000$, $\Ts=\SI{0.6}{\second}$, $d=\SI{20}{\micro\meter}$, $a=\SI{11}{\micro\meter}$, and $D=\SI{100}{\micro\meter^2/\second}$.}
\label{Fig:BER}
\end{figure*}

\subsection{Effect of Number of Emitted Molecules\label{Sec:varN}}
In Fig.~\ref{Fig:BER_Nmol}, the effect of the number of emitted molecules on the BER is shown.
If $N$ is increased, more molecules reach the receiving spheres.
Furthermore, the amplitude dependent diffusion noise is relatively getting less.
Consequently, $N$ is proportional to the signal strength.
As a result, all detection algorithms perform better, when $N$ is increased.
The simple FTD suffers from the strong ISI in the system and is not able to detect in a reasonable manner with its fixed threshold. 
The ATD, in contrast, benefits from the ISI in the system \cite{Damrath2016}.
Consequently, ATD outperforms FTD in terms of BER.
The best BER performance is achieved by MLSE, because it implies channel equalization, which counteracts ISI.
In the case of FTD, spatial diversity does not bring any enhancement compared to the SISO case for the parameters under investigation.
In contrast, repetition MIMO with EGC for ATD slightly outperforms SISO transmission in a region with few molecules.
If the power normalization of the SISO case (the emitted number of molecules in the SISO case is twice as large as in the $\txt$ MIMO case) is neglected, 
even repetition MIMO in conjunction with SD achieves a BER slightly below SISO performance.
However, the spatial diversity gain for ATD is not significant.
For MLSE, the spatial diversity gain can be more clearly observed.
The maximum BER improvement of repetition MIMO with EGC over the SISO case is by a factor of almost $10$,
yet repetition MIMO with SD shows a degradation by a factor of approximately $10$ to $10^2$.
However, neglecting the power normalization leads to a maximum improvement of almost $400$ for EGC and of approximately $10$ for SD.
Interestingly, for the system under investigation, Alamouti-type coding does not show any diversity gain.
This can be explained by the ILI in the system.
Note that ILI in repetition MIMO constructively contribute to the signal strength.
For Alamouti-type coding, however, the ILI acts competitive and thus more destructive.

\subsection{Effect of Symbol Duration\label{Sec:varTb}}
Fig.~\ref{Fig:BER_Tb} depicts the effect of symbol duration on the BER performance.
In general, an increasing symbol duration is beneficial for the communication system,
because the molecules have more time to hit the Rx during their desired symbol duration and Rx accumulates over a longer time interval.
As a result, the effect of ISI lessens and detection performance increases.
The only exception is ATD, which inherently benefits from ISI.
Therefore, in the scenario under investigation, ATD is superior to FTD for $\Ts\leq\SI{0.8}{\second}$.
For $\Ts>\SI{0.8}{\second}$, the BER of ATD increases slightly,
whereas the BER of FTD improves remarkably.
As expected, the best detection performance is achieved by MLSE.
While there is no significant gain from spatial diversity in conjunction with ATD,
repetition MIMO in conjunction with EGC and FTD outperforms the SISO case by a factor of almost $10$ at $\Ts=\SI{1.2}{\second}$.
As shown in Section~\ref{Sec:varN}, for MLSE the size of the gap between repetition MIMO with EGC and that of a SISO system is larger by a factor of almost $10$.
Furthermore, MLSE repetition MIMO with SD does not show a diversity gain, at least not for the assumed power normalization. 
As already discussed in Section~\ref{Sec:varN}, Alamouti-type coding with EGC offers no improvement over the SISO scenario.

\subsection{Effect of Transmission Distance}
In Fig.~\ref{Fig:BER_d}, the effect of the transmission distance on the system performance is shown.
In general, shorter transmission distances provide that more molecules are absorbed by Rx,
which increases the signal strength.
If the symbol duration is fixed, it also reduces the effect of ISI inside the system.
If transmission distance decreases, the BER of all detection algorithms decreases (with the exception of ATD, which inherently benefits from ISI).
For long distances ($\SI{20}{\micro\meter}\leq d \leq\SI{25}{\micro\meter}$), ATD detection performance is superior to FTD,
but for short distances ($d\leq\SI{20}{\micro\meter}$), FTD is superior to ATD.
As expected, MLSE achieves the best BER performance for all distances under consideration.
As in Section~\ref{Sec:varN} and Section~\ref{Sec:varTb}, there is no significant spatial diversity gain regarding ATD for the scenario under investigation.
Furthermore, the spatial diversity gain of repetition MIMO with EGC and MLSE is by a factor of almost $10$,
while Alamouti-type coding offers no gain at all.
In contrast to Fig.~\ref{Fig:BER_Tb}, there is also no diversity gain observed for FTD.

\subsection{Effect of Antenna Separation}
In Fig.~\ref{Fig:BER_a}, the effect of the antenna separation on the MIMO system performance is shown.
All MIMO schemes show a similar trend.
If the antenna separation is increased,
the BER is decreased.
The reason for that is in the spatial gain from the ILI,
because increasing $a$ will lead to a decreasing ILI.
Note that even for $a=\SI{17}{\micro\meter}$ there is a spatial diversity gain of repetition MIMO with EGC and MLSE.

\subsection{Effect of Diffusion Coefficient}
In Fig.~\ref{Fig:BER_D}, the effect of the diffusion coefficient on the system performance is shown.
In general, $D$ describes the mobility of a particle inside a medium.
Accordingly, $D$ has an impact on the channel impulse response.
In fact, a larger diffusion coefficient leads to a more spiky channel impulse response,
whereas a lower diffusion coefficient leads to a more flat channel impulse response.
As a result, the ISI is decreased when $D$ is increased.
Consequently, all investigated detection algorithms perform better for larger $D$.
While the spatial diversity gain from repetition MIMO with EGC for ATD increases with $D$,
it decreases with FTD.
For the scenario under investigation, the difference between SISO MLSE and repetition MIMO with EGC and MLSE is constant by a factor of approximately $10$.
As can be seen in Fig.~\ref{Fig:BER_Nmol}-\ref{Fig:BER_a}, there is no diversity gain obtained by Alamouti-type coding.

\section{Conclusion\label{Sec:Conclusion}}
In this paper, we have presented a diffusion-based molecular $\txt$ MIMO communication system in a 3-D environment.
Channel coefficients were obtained from a trained ANN and incorporated into performance evaluations.
Motivated from the potential of spatial diversity in classical wireless communication,
this paper introduced different spatial diversity algorithms to the area of MC and analyzed their performances.
An the transmitter side, Alamouti-type coding and repetition MIMO coding were proposed.
An the receiver side, selection diversity, equal-gain combining, and maximum-ratio combining were presented as receiver combining strategies.
In addition, fixed threshold detection, adaptive threshold detection and maximum-likelihood sequence estimation were adapted to the $\txt$ MIMO scenario.

The diversity gain was studied by means of BER simulations,
where different system parameters were varied to show the effect on the system performance.
Similar to the SISO case, MLSE outperforms FTD and ATD, while ATD outperforms FTD when more ISI is present, since ATD profits from ISI.
Furthermore, FTD and MLSE performance benefit from a higher number of emitted molecules, a larger symbol duration, a shorter transmission distance and a higher diffusion coefficient.
A significant spatial diversity gain can only be achieved by repetition MIMO with EGC and MLSE for the scenario under investigation.
In contrast, Alamouti-type coding fails to show a practical performance in the context of MC.
Even in conjunction with MLSE, it suffers from the discrepancy between averaged and actual channel coefficients.
Spatial diversity gain is dependent on the antennas separation distance.
The simulation showed that the best BER performance is achieved, when the receive and transmit antennas are as close together as possible.

Future work will include the realization of spatial diversity in a practical system, 
analyzing spatial diversity for unsymmetrical cases,
and expanding the system to a higher number of transmit and/or receive antennas.




%

\bibliographystyle{IEEEtran}
\bibliography{mybibs_MIMO_diversity}

\begin{IEEEbiography}{Martin Damrath} (S'16) received the M.Sc. degree
in electrical engineering from the University of Kiel,
Germany, in 2014. Since 2014, he is working as
a research and teaching assistant in the Information
and Coding Theory Lab towards his Dr.-Ing.
(Ph.D.) degree in the Faculty of Engineering,
University of Kiel. His current research interests
include high-order modulation schemes, superposition
modulation, and diffusion-based molecular
communication.
\end{IEEEbiography}

\begin{IEEEbiography}{H. Birkan Yilmaz} (S'10-M'12)
	received his M.Sc. and Ph.D. degrees in Computer Engineering from Bogazici University in 2006 and 2012, respectively, his B.S. degree in Mathematics in 2002. Currently, he works as a post-doctoral researcher at Yonsei Institute of Convergence Technology, Yonsei University, Korea. He was awarded TUBITAK National Ph.D. Scholarship during his Ph.D. studies. He was the co-recipient of the best demo award in IEEE INFOCOM (2015) and best paper award in ISCC (2012) and AICT (2010). He is a member of TMD (Turkish Mathematical Society). His research interests include cognitive radio, spectrum sensing, molecular communications, and detection and estimation theory.
\end{IEEEbiography}

\begin{IEEEbiography}{Chan-Byoung Chae}(SM'12)
is the Underwood Distinguished Professor in the School of Integrated Technology, Yonsei University. Before joining Yonsei University, he was with Bell Labs, Alcatel-Lucent, Murray Hill, New Jersey from 2009 to 2011, as a member of technical staff, and Harvard University, Cambridge, Massachusetts from 2008 to 2009, as a postdoctoral research fellow. He received his Ph.D. degree in electrical and computer engineering from The University of Texas at Austin in 2008, where he was a member of the Wireless Networking and Communications Group (WNCG). Prior to joining UT, he was a research engineer at the Telecommunications R\&D Center, Samsung Electronics, Suwon, Korea, from 2001 to 2005. He has served/serves as an Editor for the \emph{IEEE Communications Magazine} (2016-present), the \emph{IEEE Trans. on Wireless Communications (2012~present)}, the \emph{IEEE Trans. on Molecular, Biological, and Multi-scale Comm. (2015-present)}, the \emph{IEEE Wireless Communications Letters (2016-present)}, and the \emph{IEEE/KICS Jour. of Comm. Networks (2012-present)}.

He was the recipient/co-recipient of the Underwood Distinguished Professor Award (2016) and the Outstanding Teaching Award (2016) from Yonsei University, the Yonam Research Award from LG Yonam Foundation (2016), the Outstanding Professor Award from IITP (2016), the Best Young Professor Award from the College of Engineering, Yonsei University (2015), the IEEE INFOCOM Best Demo Award (2015), the IEIE/IEEE Joint Award for Young IT Engineer of the Year (2014), the KICS Haedong Young Scholar Award (2013), the IEEE Signal Processing Magazine Best Paper Award (2013), the IEEE ComSoc AP Outstanding Young Researcher Award (2012), the IEEE VTS Dan. E. Noble Fellowship Award (2008), two Gold Prizes (1st) in the 14th/19th Humantech Paper Contest. He also received the Korea Government Fellowship (KOSEF) during his Ph.D. studies.
\end{IEEEbiography}

\begin{IEEEbiography}{Peter Adam Hoeher}(F'14) received Dipl.-Ing.
(M.Sc.) and Dr.-Ing. (Ph.D.) degrees in Electrical
Engineering from the RWTH Aachen University,
Aachen, Germany, and the University of Kaiserslautern,
Kaiserslautern, Germany, in 1986 and
1990, respectively. From October 1986 to September
1998, he was with the German Aerospace Center
(DLR), Oberpfaffenhofen, Germany. From December
1991 to November 1992 he was on leave
at AT\&T Bell Laboratories, Murray Hill, NJ, USA.
In October 1998 he joined the University of Kiel,
Germany, where he is a full professor of electrical and information engineering.
His research interests are in the general area of communication theory and
applied information theory with applications in wireless communications and
underwater communications, including advanced digital modulation techniques,
channel coding, iterative processing, equalization, multiuser detection,
interference cancellation, channel estimation, and joint communication and
navigation. Dr. Hoeher received the Hugo-Denkmeier-Award '90 and the ITG
Award ’07. Between 1999 and 2006 he served as an Associated Editor for \emph{IEEE
Trans. on Communications}. Since 2014 he is an IEEE Fellow ``for
contributions to decoding and detection that include reliability information."
\end{IEEEbiography}
\end{document}